\documentclass[aps,prd,nofootinbib,twocolumn,superscriptaddress,floatfix]{revtex4}
\usepackage{mathrsfs}
\usepackage{latexsym}
\usepackage{amsmath}
\usepackage{amssymb}
\usepackage{graphicx}
\usepackage{longtable}
\usepackage{multirow}
\usepackage{bbm}
\usepackage{epsfig}
\usepackage[usenames]{color}
\usepackage[colorlinks=true,citecolor=blue,linkcolor=blue]{hyperref}
\allowdisplaybreaks

\begin{document}

%\begin{CJK*}{GBK}{song}

%\begin{CJK*}{GB}{}

\title{Quark matter with an anisotropic momentum distribution}

\author{Wei-bo He} %\author{Guo-yun Shao}   \author{Xue-yan Gao} \author{Chong-long Xie} \author{Zhong-lv Huang} 
\affiliation{ MOE Key Laboratory for Non-equilibrium Synthesis and Modulation of Condensed Matter, 
	School of Physics, Xi’an Jiaotong University, Xi’an 710049, China}
\affiliation{ School of Physics, Peking University, Beijing, 100871, China}
\author{Guo-yun Shao}   
\email[Corresponding author: ]{gyshao@mail.xjtu.edu.cn} 
%%\thanks{These authors contributed equally to this work}
\affiliation{ MOE Key Laboratory for Non-equilibrium Synthesis and Modulation of Condensed Matter, 
School of Physics, Xi’an Jiaotong University, Xi’an 710049, China}

%\author{Xin-ran Yang}   
%\affiliation{ MOE Key Laboratory for Non-equilibrium Synthesis and Modulation of Condensed Matter, School of Physics, Xi’an Jiaotong University, Xi’an 710049, China}

%\author{Chong-long Xie}
%\affiliation{ MOE Key Laboratory for Non-equilibrium Synthesis and Modulation of Condensed Matter, School of Physics, Xi’an Jiaotong University, Xi’an 710049, China}

%\author{Xue-yan Gao}
%\affiliation{ MOE Key Laboratory for Non-equilibrium Synthesis and Modulation of Condensed Matter, School of Physics, Xi’an Jiaotong University, Xi’an 710049, China}

\begin{abstract}
Motivated by the anisotropic momentum distribution of particles in heavy-ion collisions, we study the
angular dependence of quark average momentum and quark distribution function in the Polyakov-Nambu--Jona-Lasinio (PNJL) quark model.
We also investigate the phase transitions and net baryon number fluctuations in anisotropic quark matter. The numerical results suggest that the QCD phase structure and isentropic trajectories are sensitive to the anisotropic parameter at finite density, in particular, in the area near the critical region and the first-order phase transition. Compared with the isotropic quark matter, the values of baryon number kurtosis and skewness at lower collision energies are possibly enhanced  with the anisotropic momentum distribution  squeezed along the  direction of nucleus-nucleus collision in experiments.

\end{abstract}

%\pacs{12.38.Mh, 25.75.Nq}

\maketitle
\section{introduction} 

Exploring the phase structure of  quantum chromodynamics~(QCD)  is an important topic in both  theoretical and experimental nuclear physics.
The calculation from lattice QCD (LQCD) indicates that the transformation from quark-gluon plasma~(QGP) to hadrons  is a smooth crossover~\cite{Aoki06, Gupta11, Bazavov12, Borsanyi13,Bazavov14, Bazavov17,Borsanyi14,Borsanyi20}  at vanishing and small baryon chemical potentials. Many QCD inspired models/approaches further predict that there exists a first-order phase transition with a critical endpoint (CEP) connecting with a crossover  transformation at finite temperature and chemical potential~(e.g.,\cite{Fukushima04,Ratti06, Meisinger96, Costa10, Sasaki12, Ferreira14, Schaefer10, Skokov11, Qin11,Gao16, Fischer14, Maslov23, Fu20, Shao2018, Liu2018}).
The  QCD phase structure  can be probed due to  the energy-dependent behavior of the ratios of net-baryon number fluctuations at chemical freeze out~\cite{Stephanov}. The cumulants of net proton~(proxy for baryon) have been measured in the Beam Energy Scan experiments at RHIC STAR, and the nonmonotonic energy dependence of  the net-proton number fluctuations have been discovered ~\cite{Luo2017, Adam21, STAR23}.  The experimental data  has aroused a wide discussion about  whether the QCD critical region has been reached. %More accurate measurement of BES-II and the relevant experimental projects at NICA/FAIR/J-PARC/HIAF will provide us more information about the QCD phase diagram.

The isotropic momentum distribution is usually assumed to explore the phase structure in LQCD simulations and QCD inspired models. However, 
deviations from perfect isotropy are expected for a real quark-gluon plasma. A large momentum-space anisotropy can arise  due to the rapid longitudinal expansion of the matter created in relativistic heavy ion collisions and the anisotropy possibly survive during the entire evolution of the medium~\cite{Dumitru09}. A similar result exists in hydrodynamic simulations. %For a fireball created in heavy-ion collision~(HIC) experiments, 
The ideal relativistic hydrodynamics predict that the QGP would  tends to be isotropic on a timescale $\tau \sim 0.5\, \mathrm{fm} /\mathrm{c}$~\cite{Huovinen01, Hirano02}.
In practice, however, with the inclusion of viscous correction  sizable differences between the transverse and longitudinal pressure can still be observed at times $\tau \lesssim 2$ fm/c~\cite{Muronga02,Heinz06,Romatschke07,Song08,Denicol10,Calzetta15,Denicol14,Jaiswal14}. %associated with the existence of a nonequilibrium hydrodynamic attractor. 
%Moreover, the anisotropy of pressure  increases at the transverse/longitudinal
%edges of the quark-gluon plasma~(QGP). 
Recently, the anisotropic hydrodynamics has also been developed to account  for large deviations from isotropy in momentum space, which provides a more accurate description of non-equilibrium dynamics than usual relativistic hydrodynamic models~\cite{Tinti14, Alqahtani18, Alqahtani17,Almaalol19, Alalawi20}. 

A natural question aroused is how the anisotropic momentum distribution affects the QCD phase transition and final observables in experiments. To answer this question, it is necessary to explore various properties of anisotropic QGP.
The  anisotropic distribution of particles in momentum space  can be phenomenologically described by introducing a spheroidally anisotropic distribution function by stretching or squeezing the isotropic distribution along  one of the directions. In Ref.~\cite{Romatschke03}, a popular one-particle distribution function in anisotropic momentum space was firstly proposed in the pioneer work  by Romatschke and Strickland, %$f^{\operatorname{aniso}}(\mathbf{p})=\frac{1}{\texttt{exp} \left[\left(\sqrt{\mathbf{p}^{2}+\xi(\mathbf{p} \cdot \mathbf{n})^{2}+m^{2}}-\mu_{}^{}\right) / T^{}\right]+1}$
 $f^{\operatorname{aniso}}(\mathbf{p})=\left[\texttt{exp} \left[\left(\sqrt{\mathbf{p}^{2}+\xi(\mathbf{p} \cdot \mathbf{n})^{2}+m^{2}}-\mu_{}^{}\right) / T^{}\right] \pm1\right]^{-1}$, where $\xi$ is a parameter indicating the strength and type of momentum-space anisotropy and $ \mathbf{n}$ is the anisotropy direction. Such a parametrization is interesting in heavy-ion collisions with the parton distribution to be squeezed along the beam direction. 
By far, the anisotropic distribution in momentum space has been considered to study various issues, such as,  collective modes~\cite{Carrington21},  Quarkonium states~\cite{Dumitru092, Boguslavski21},  photon and dilepton production~\cite{Hauksson21,Kasmaei20},  as well as transport coefficients~\cite{Thakur17,Rath19}.
%The leading-anisotropic correction subsequently affects the suppression of quarkonium production at RHIC and LHC. 
 
In this work, we focus on exploring the phase transition in anisotropic QCD medium. Since the fluctuations of conserved charges are closely related to QCD phase transitions~\cite{Shao2018,Shao2020,Fu21} and are also the sensitive probes to diagnose the QCD phase structure in HIC experiments~\cite{Stephanov, Luo2017, Adam21, STAR23}, we further explore the correlation of baryon number fluctuations with the phase transitions in anisotropic QCD matter. The potential impact on experimental data analysis of baryon number fluctuations are also discussed.
The 2+1 flavor PNJL quark model is taken in the calculation, which describe well both the chiral phase transition and (de)confinement phase transition of QCD.  
%The numerical results indicate that both the baryon number fluctuations and the first-order phase transition at low temperature are sensitive to the anisotropic parameter $\xi$. 

\section{formulas for quark matter with an anisotropic momentum distribution}
We first simply introduce the thermodynamic formulas for isotropic quark matter, and then extend the relevant formulas to the case of anisotropy in momentum space. 
The Lagrangian density in the 2+1 flavor PNJL model is given by~\cite{Shao2018,he2022}
\begin{eqnarray}
	\mathcal{L}&\!=&\!\bar{q}(i\gamma^{\mu}D_{\mu}\!+\!\gamma_0\hat{\mu}\!-\!\hat{m}_{0})q\!+\!
	G\sum_{k=0}^{8}\big[(\bar{q}\lambda_{k}q)^{2}\!+\!
	(\bar{q}i\gamma_{5}\lambda_{k}q)^{2}\big]\nonumber \\
	&&-K\big[\texttt{det}_{f}(\bar{q}(1+\gamma_{5})q)+\texttt{det}_{f}
	(\bar{q}(1-\gamma_{5})q)\big]\nonumber \\ \nonumber \\
	&&-U(\Phi[A],\bar{\Phi}[A],T),
\end{eqnarray}
where $q$ denotes the quark fields with three flavors, $u,\ d$, and
$s$; $\hat{m}_{0}=\texttt{diag}(m_{u},\ m_{d},\
m_{s})$ in flavor space; $G$ and $K$ are the four-point and
six-point interacting constants, respectively.  The $\hat{\mu}=diag(\mu_u,\mu_d,\mu_s)$ are the quark chemical potentials.% which are related to chemical potentials of the conserved charges through the Eq.~(\ref{chemical_1}) and (\ref{chemical_2}).

The covariant derivative in the Lagrangian is defined as $D_\mu=\partial_\mu-iA_\mu$.
The gluon background field $A_\mu=\delta_\mu^0A_0$ is supposed to be homogeneous
and static, with  $A_0=g\mathcal{A}_0^\alpha \frac{\lambda^\alpha}{2}$, where
$\frac{\lambda^\alpha}{2}$ is $SU(3)$ color generators.
The effective potential $U(\Phi[A],\bar{\Phi}[A],T)$ is expressed with the traced Polyakov loop
$\Phi=(\mathrm{Tr}_c L)/N_C$ and its conjugate
$\bar{\Phi}=(\mathrm{Tr}_c L^\dag)/N_C$. The Polyakov loop $L$  is a matrix in color space
\begin{equation}
	L(\vec{x})=\mathcal{P} exp\bigg[i\int_0^\beta d\tau A_4 (\vec{x},\tau)   \bigg],
\end{equation}
where $\beta=1/T$ is the inverse of temperature and $A_4=iA_0$. 

In the mean field approximation, the constituent quark mass can be derived as
\begin{equation}
	M_{i}=m_{i}-4G\phi_i+2K\phi_j\phi_k\ \ \ \ \ \ (i\neq j\neq k),
	\label{mass}
\end{equation}
where $\phi_i$ stands for quark condensate of the flavor $i$.
The thermodynamical potential of bulk quark matter is derived as \cite{Fukushima:2003fw, Meisinger:1995kp, Ghosh:2007wy, Mukherjee:2006hq, Roessner:2006xn,Ghosh06}
%\begin{widetext}
\begin{eqnarray}
	\Omega&=&U(\bar{\Phi}, \Phi, T)+2G\left({\phi_{u}}^{2}
	+{\phi_{d}}^{2}+{\phi_{s}}^{2}\right)-4K\phi_{u}\,\phi_{d}\,\phi_{s} \nonumber\\
	&&-2\int_\Lambda \frac{\mathrm{d}^{3}p}{(2\pi)^{3}}3(E_u+E_d+E_s) \nonumber \\
	&&-2T \sum_{i=u,d,s}\int \frac{\mathrm{d}^{3}p}{(2\pi)^{3}} (\mathrm{ln}\mathcal{A}_1+\mathrm{ln}\mathcal{A}_2),
\end{eqnarray}
where $\mathcal{A}_1=1
+3\Phi e^{-(E_i-\mu_i)/T}+3\bar{\Phi} e^{-2(E_i-\mu_i)/T}+e^{-3(E_i-\mu_i)/T}$,  $\mathcal{A}_2=1+3\bar{\Phi} e^{-(E_i+\mu_i)/T}
+3\Phi e^{-2(E_i+\mu_i)/T}+e^{-3(E_i+\mu_i)/T} $, and $\mu_i$ is the quark chemical potential. $E_i=\sqrt{\mathbf{p}^{\,2}+M_i^2}$ is the dispersion relation of quark in isotropic QCD medium.

The Polyakov-loop effective potential \cite{Roessner:2006xn} taken in this study   is
\begin{eqnarray} \label{U}
	\frac{U(\Phi,\bar{\Phi},T)}{T^4}&=&-\frac{a(T)}{2}\bar{\Phi}\Phi +b(T)\,\mathrm{ln}\big[1-6\bar{\Phi}\Phi\\ \nonumber
	&&+4(\bar{\Phi}^3+\Phi^3)-3(\bar{\Phi}\Phi)^2\big],
\end{eqnarray}
where
\begin{equation}
	\!a(T)\!=\!a_0\!+\!a_1\big(\frac{T_0}{T}\big)\!+\!a_2\big(\frac{T_0}{T}\big)^2 \,\,\,\texttt{and}\,\,\,\,\, b(T)\!=\!b_3\big(\frac{T_0}{T}\big)^3.
\end{equation}
The logarithmic effective potential in Eq.~(\ref{U}) effectively includes the Vandermonde term  from the Jacobian of transformation from Wilson line to Polyakov loop, and it rectifies the anomaly of traced Polyakov-loop with $\Phi>1$ at high temperature for a simple polynomial form of $U(\Phi,\bar{\Phi},T)$~\cite{Ratti06}. 

The parameters $a_i$, $b_i$ listed in Table. \ref{tab:1} are fitted according to the lattice simulation of  QCD thermodynamics in
pure gauge sector, %And $T_0$ is found to be 270 MeV as the critical temperature for the deconfinement phase transition of gluon part at zero chemical potential~\cite{Fukugita90}. 
and $T_0=210$\, MeV
is implemented in the calculation.  In the numerical calculation, a cut-off $\Lambda$ is implemented in 3-momentum
space for divergent integrations. We take the model parameters obtained in~\cite{{Rehberg:1995kh}}:
$\Lambda=602.3$ MeV, $G\Lambda^{2}=1.835$, $K\Lambda^{5}=12.36$,
$m_{u,d}=5.5$  and $m_{s}=140.7$ MeV, determined
by fitting $f_{\pi}=92.4$ MeV,  $M_{\pi}=135.0$ MeV, $m_{K}=497.7$ MeV and $m_{\eta}=957.8$ MeV. 
\begin{table}[ht]
	\centering
	\caption{Parameters in the Polyakov-loop potential~\cite{ Roessner:2006xn}}
	\label{tab:1}
	\begin{tabular*}{\columnwidth}{@{\extracolsep{\fill}}llll@{}}
		\hline
		\multicolumn{1}{@{}l}{$a_0$} & $a_1$ & $a_2$ & $b_3$\\
		\hline
		$ 3.51$                   & -2.47        &  15.2      & -1.75               \\ 
		\hline
	\end{tabular*}
\end{table}

By minimizing the thermodynamical potential
\begin{equation}
\frac{\partial\Omega}{\partial\phi_u}=\frac{\partial\Omega}{\partial\phi_d}=\frac{\partial\Omega}{\partial\phi_s}=\frac{\partial\Omega}
{\partial\Phi}=\frac{\partial\Omega}{\partial\bar\Phi}=0.
\end{equation}
we can derive the equations of motion in medium as
\begin{equation}\label{cond}
\phi_i\!=\!-\!2N_{c}\!\int\frac{d^{3} p}{(2\pi)^{3}}\frac{M_i}{E_i}
\big(1-f_i(p)-\bar{f}_i(p)\big) \,(i=u,d,s),
\end{equation}
\begin{equation}
\!\frac{\partial U}{\partial \Phi}-6T\!\sum_{i=u d s}\!\int\!\frac{d^{3} p}{(2\pi)^{3}}\!\left(\frac{1}{\mathcal{A}_1}e^{-\frac{\!\!E_{i}\!-\!\mu_{i}}{T} }\!+\!\frac{1}{\mathcal{A}_2}e^{\!-2\frac{\!E_{i}\!+\!\mu_{i}}{T} }\!\right)\!=\!0,\!
\end{equation}
and
\begin{equation}\label{barPhi}
\!\frac{\partial U}{\partial \bar \Phi}\!-\!6T\!\sum_{i=u d s}\!\int\!\frac{d^{3} p}{(2\pi)^{3}}\!\left(\!\frac{1}{\mathcal{A}_1}e^{-2\frac{\!\!E_{i}\!-\!\mu_{i}}{T} }\!+\!\frac{1}{\mathcal{A}_2}e^{\!-\frac{\!E_{i}\!+\!\mu_{i}}{T} }\!\right)\!=\!0.\!
\end{equation}
In Eq.~(\ref{cond}), 
\begin{equation}\label{distribution}
 \!\!\!f_{i}(p)\!=\!\frac{\!\Phi e^{\!-\!(\!E_i\!-\!\mu_i\!)\!/\!T}\!+\!2\bar{\Phi} e^{\!-\!2(\!E_i\!-\!\mu_i\!)\!/\!T}+e^{-3(E_i-\mu_i)/T}}
  {\!1\!+\!3\Phi e^{\!-\!(\!E_i\!-\!\mu_i\!)\!/\!T}\!+\!3\bar{\Phi} e^{\!-\!2(\!E_i\!-\!\mu_i)/T}\!+\!e^{\!-\!3(E_i\!-\!\mu_i)\!/\!T}}
\end{equation}
and
\begin{equation}\label{antif}
  \!\!\bar f_{i}(p)\!=\!\frac{\! \bar \Phi e^{\!-\!(\!E_i\!+\!\mu_i\!)\!/\!T}\!+\!2\Phi e^{\!-\!2(\!E_i\!+\!\mu_i\!)\!/\!T}+e^{-3(E_i+\mu_i)/T}}
  {\!1\!+\!3\bar \Phi e^{\!-\!(\!E_i\!+\!\mu_i\!)\!/\!T}\!+\!3\Phi e^{\!-\!2(\!E_i\!+\!\mu_i\!)/T}\!+\!e^{\!-\!3(\!E_i\!+\!\mu_i)\!/\!T}} 
\end{equation}
are modified Fermion distribution functions of quark and antiquark, respectively. For a given temperature and chemical potential, the values of order parameters, $\phi_u, \phi_d, \phi_s, \Phi$ and $\bar{\Phi}$ can be derived by solving Eqs.~(\ref{cond}-\ref{barPhi}).

For anisotropic quark matter,  the dispersion relation of quasiparticles needs to be modified according to the anisotropic momentum distribution. In this study we take the Romatschke and Strickland~(RS) scheme~\cite{Romatschke03} in which the  nontrivial dispersion relation for a particle with mass $m$ is described by
\begin{equation}
E^{\operatorname{aniso}}=\sqrt{\mathbf{p}^{2}+\xi(\mathbf{p} \cdot \mathbf{n})^{2}+m^{2}}.
\end{equation}
Correspondingly, the distribution function of bosons and fermions at temperature $T$ and chemical potential $\mu$ is 
\begin{equation}
\!f^{\operatorname{aniso}}(\mathbf{p})=\frac{1}{\!\left[\texttt{exp} \!\left(\!\sqrt{\mathbf{p}^{2}\!+\!\xi(\mathbf{p} \cdot \mathbf{n})^{2}\!+\!m^{2}}\!-\!\mu_{}^{}\!\right)\! / T^{}\right]\! \pm1}\!,\!
\end{equation}
where $ \mathbf{n}$ is the unit vector along the anisotropy direction. 
The anisotropic parameter $\xi$ is defined as 
\begin{equation}
\xi=\frac{\left\langle p_{\perp}^{2}\right\rangle} {2\left\langle p_{\|}^{2}\right\rangle}-1,
\end{equation}
where $p_{\|}=\mathbf{p} \cdot \mathbf{n}$ is the momentum component parallel to the direction $\mathbf{n}$, and $p_{\perp}=|\mathbf{p}-(\mathbf{p} \cdot \mathbf{n}) \cdot \mathbf{n}|$ is the  component perpendicular to $\mathbf{n}$. The range of  $\xi$ is $-1<0<\infty$, which indicates the strength and type of momentum-space anisotropy. The momentum distribution is isotropic for $\xi=0$. For  $\xi>0$, the momentum distribution is squeezed along  $\mathbf{n}$ direction. For  $-1<\xi<0$, it corresponds to a stretched momentum distribution along the direction  $\mathbf{n}$.

Such a parametrization is interesting in heavy-ion collisions with the momentum distribution to be squeezed or stretched along one direction. 
Combined with the heavy-ion collision experiments, it is reasonable and convenient to choose $\mathbf{n}$ along the direction of nucleon-nucleon collision. With this parametrization the dispersion relation for quarks and antiquarks becomes 
\begin{equation}\label{newdisper}
E_i^{\operatorname{aniso}}=\sqrt{{p}^{2}+\xi{p}^2  \cos\theta^{2}+M_i^{2}}.
\end{equation} 
The dispersion relation $E_i=\sqrt{\mathbf{p}^{\,2}+M_i^2}$ in isotropic matter in Eqs.~(\ref{mass})-(\ref{antif}) will be replaced with the new one to explore the properties of anisotropic quark matter. At the same time, all the relevant three dimensional integrations will be performed with the consideration of anisotropic momentum distribution.
We assume that the gluon field is a background field, so the anisotropic distribution of quark momentum does not affect the form of $U(\Phi, \bar{\Phi}, T)$. The modeling of anisotropic medium primarily relies on a quasiparticle description where the medium effects are encoded in the effective distribution functions.

\section{numerical results and discussions}
\begin{figure} [htbp]% figuur 1
\begin{minipage}{\columnwidth}
\centering
\includegraphics[scale=0.4]{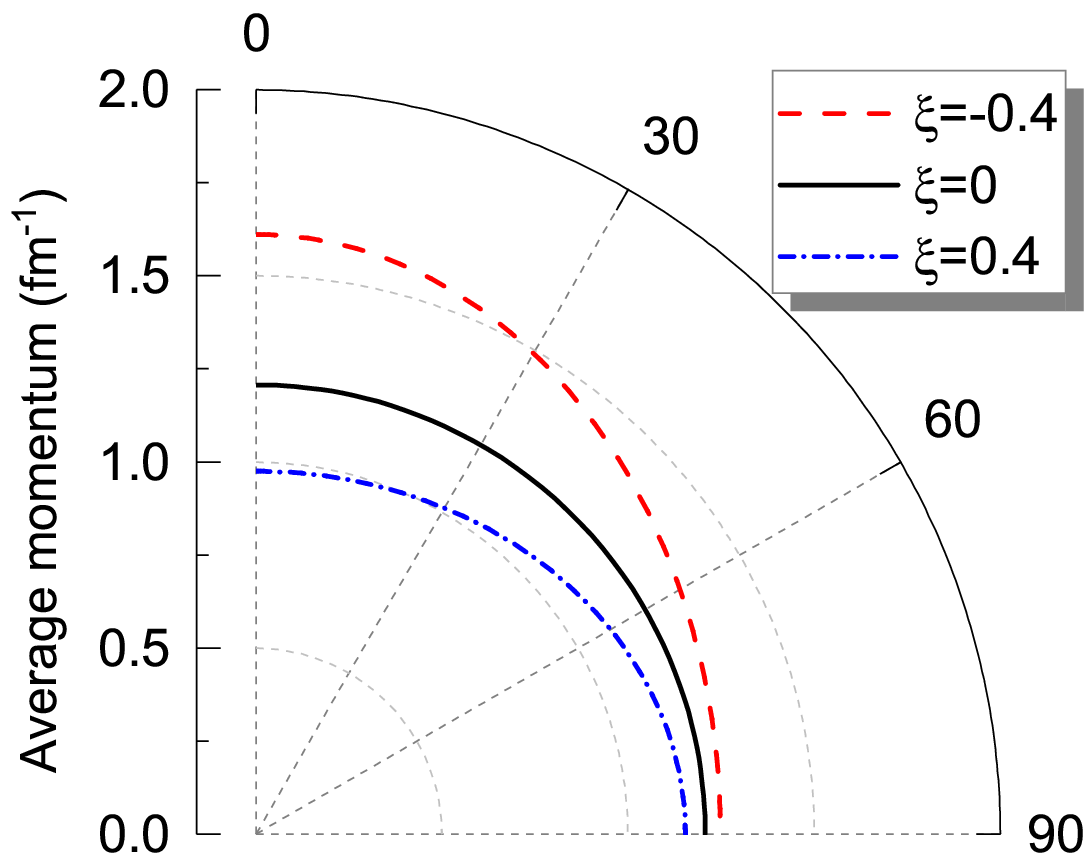}
\end{minipage}
\caption{Angular dependence of average momentum of $u(d)$ quark for $\xi=-0.4, 0, 0.4$ at $T=180\,$MeV and $\mu_B=0$. Assuming $\theta=0$ is the anisotropic direction $\texttt{n}$.}
\label{fig:1}
\end{figure}

To understand the anisotropic distribution of momentum under different parameter $\xi$ with a dynamical quark mass, we first plot in Fig.~\ref{fig:1} the angular dependence of the average momentum of $u(d)$ quark for $\xi=-0.4, 0, 0.4$, respectively. The  mean value of momentum at  the angle $\theta$ is defined as
\begin{equation}
\bar{p}(\theta,\xi)=\frac{\int_{0}^{\infty} p f(p, \theta,\xi) d p}{\int_{0}^{\infty} f(p, \theta,\xi) d p}, 
\end{equation} 
where the quark distribution function $f(p, \theta,\xi)$ is given in Eq.~(\ref{distribution}) with the modified dispersion relation in Eq.~(\ref{newdisper}).  The function $f(p, \theta, \xi)$ also depends on the order parameters~($\phi_i, \Phi, \bar{\Phi}$), which are determined for a given $T$ and $ \mu_B$ by solving Eqs.~(\ref{cond})-(\ref{barPhi}). As an example, the  average of angular dependent momentum  shown in Fig.~\ref{fig:1} is calculated at $T=180\,$MeV and $\mu_B=0$.

In the case  of $\xi=0$, Fig.~\ref{fig:1} shows that  $\bar p(\theta,\xi)$ is independent of angle $\theta$, which is just the feature of momentum isotropy.
In the case of $\xi=-0.4$, the average of momentum is anisotropic with a maximum value at $\theta=0$ (i.e., along the anisotropic direction $\texttt{n}$) and a minimum value at $90^\circ$, perpendicular to the anisotropic direction  $\texttt{n}$. The opposite happens in the case of $\xi=0.4$. %{\it However, note that the angular dependence of  $\bar p$ for $\xi=0.4$ cannot be simply derived  by a simple rotation from the data of $\xi=-0.4$.}
With the assumption of  $\theta=0$ being the anisotropic direction $\texttt{n}$, Fig.~\ref{fig:1} clearly demonstrates that the anisotropic momentum is stretched along the anisotropic direction $\texttt{n}$ for  $\xi<0$ and squeezed for $\xi>0$. %This figure also shows that the 

\begin{figure} [htbp]% figuur 1
\begin{minipage}{\columnwidth}
\centering
\includegraphics[scale=0.4]{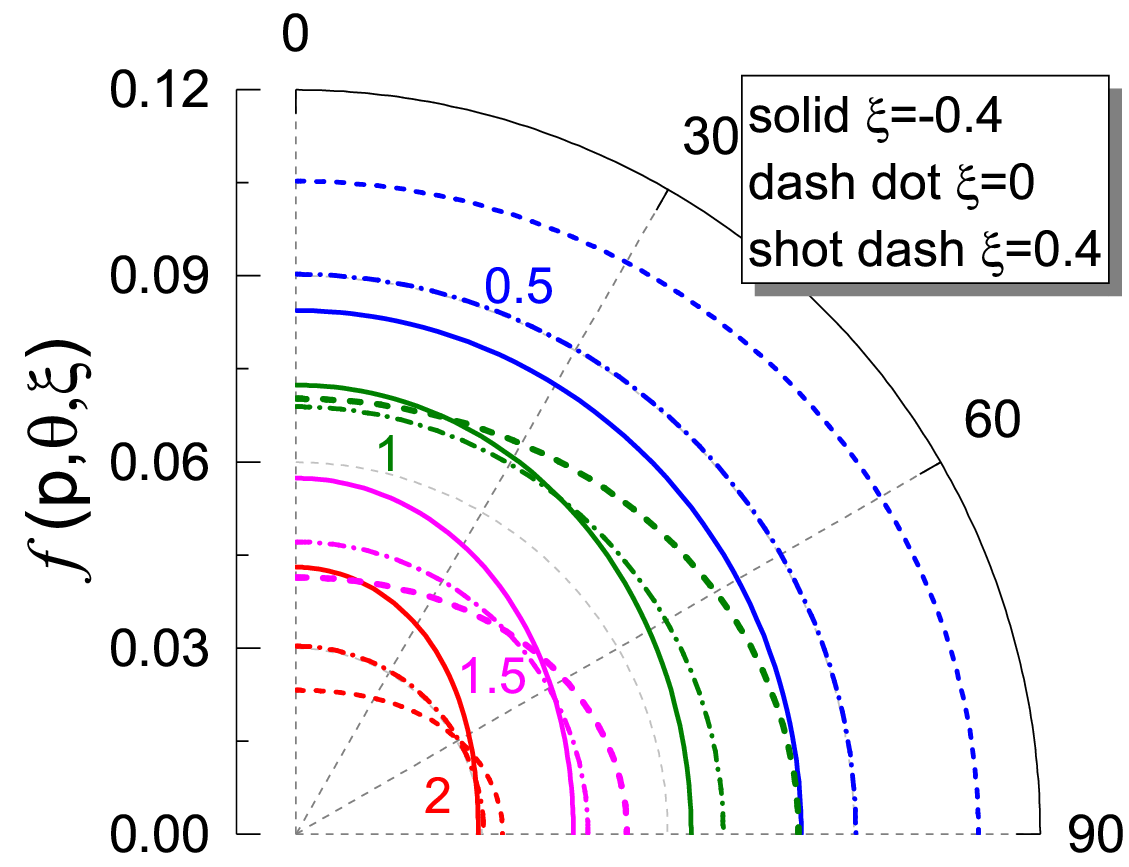}
\end{minipage}
\caption{ Angular  dependence of quark distribution function $f(p, \theta, \xi)$ for  $\xi=-0.4, 0, 0.4$ with several fixed momentums, $p=0.5, 1, 1.5, 2\,$ fm$^{-1}$ at $T=180\,$MeV and $\mu_B=0$. $\theta=0$ is the anisotropic direction.}
\label{fig:2}
\end{figure}

In Fig.~{\ref{fig:2}}, we further present the angular dependence of quark distribution function $f(p, \theta, \xi)$ for different $\xi$ with several fixed momentums, $p=0.5, 1, 1.5, 2\,$fm$^{-1}$. This figure shows that  $\left.f(p, \theta, \xi)\right|_{p}$ descends~(increases) as $\theta$ increases from $0^{\circ}$ to $90^{\circ}$ for $\xi=-0.4$ ($\xi=0.4$). 
It also indicates that, at a fixed angle, the anisotropy of $f(p, \theta, \xi)$ changes for different $\xi$ at the same momentum $p$. This possibly provides a potential method to test the anisotropic parameter $\xi$ if the angular and momentum dependence of particle distribution in HIC experiments is available. A further study will be carried on this issue.

\begin{figure} [htbp]% figuur 1
\begin{minipage}{\columnwidth}
\centering
\includegraphics[scale=0.37]{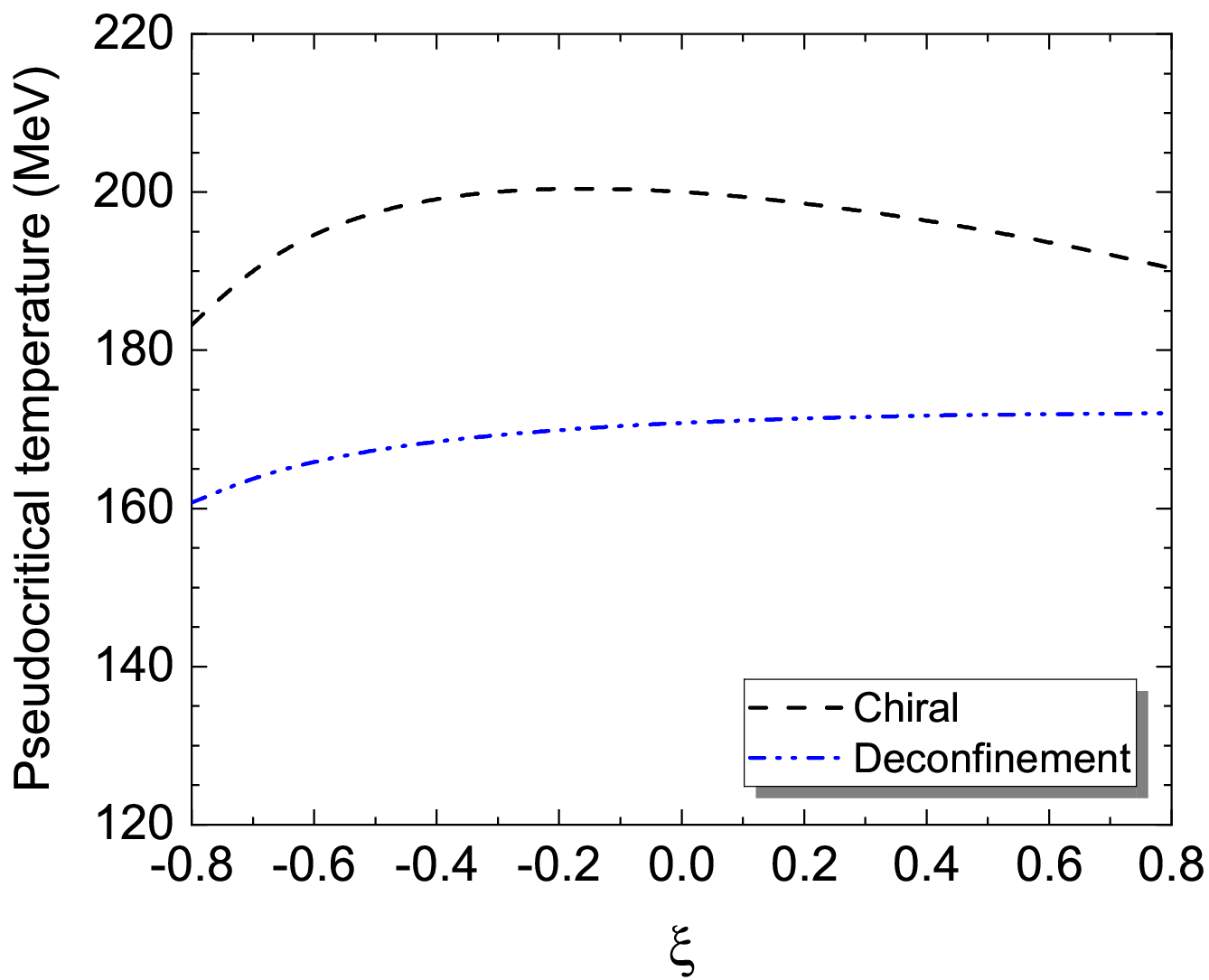}
\end{minipage}
\caption{ Pseudocritical critical temperatures of the chiral and (de)confinement  transformation as functions of anisotropic parameter $\xi$.}
\label{fig:3}
\end{figure}

Now we investigate the QCD phase transition at vanishing chemical potential with different anisotropic parameter $\xi$. We plot in Fig.~\ref{fig:3} the pseudocritical critical temperatures of both the chiral and deconfinement  transformation derived at $\mu_B=0$ as functions of anisotropic parameter $\xi$. For each value of $\xi$, the pseudocritical temperature of chiral phase transition ($T_\chi$) is derived at the location where  ($\partial {\phi_u} / \partial T)_{\mu_B=0}$ takes the maximum. Similarly, the pseudocritical temperature of deconfinement transformation ($T_D$)  is derived with the condition of ($\partial {\Phi} / \partial T)_{\mu_B}=0$ taking the maximum.
Fig.~\ref{fig:3} shows  that the  pseudocritical temperature of color deconfinement increases monotonically with the increase of $\xi$. However,  the pseudocritical temperature  of chiral crossover changes nonmonotonically with the variation of $\xi$. The maximum value of $T_\chi$ appear at $\xi=-0.16$. %$T_\chi$ decreases with the enhance of anisotropy with much larger $|\xi|$.
 Note also that $T_\chi(\xi)$ is  not simply symmetrical on both sides of $\xi=0$.  
 
Note that in this version of PNJL model the locations of chiral and deconfinement transitions do not coincide at vanishing chemical potential. Some extensions, such as considering the entanglement interaction between the chiral condensate and the Polyakov loop (EPNJL) or the eight-quark interaction, can reduce the difference between them to some extent. 
In this study we temporarily ignore these factor, and focus on the different effect of anisotropic momentum distribution on the pseudocritical temperatures of the two phase transitions with different $\xi$ in this model.

\begin{figure} [htbp]% figuur 1
\begin{minipage}{\columnwidth}
\centering
\includegraphics[scale=0.62]{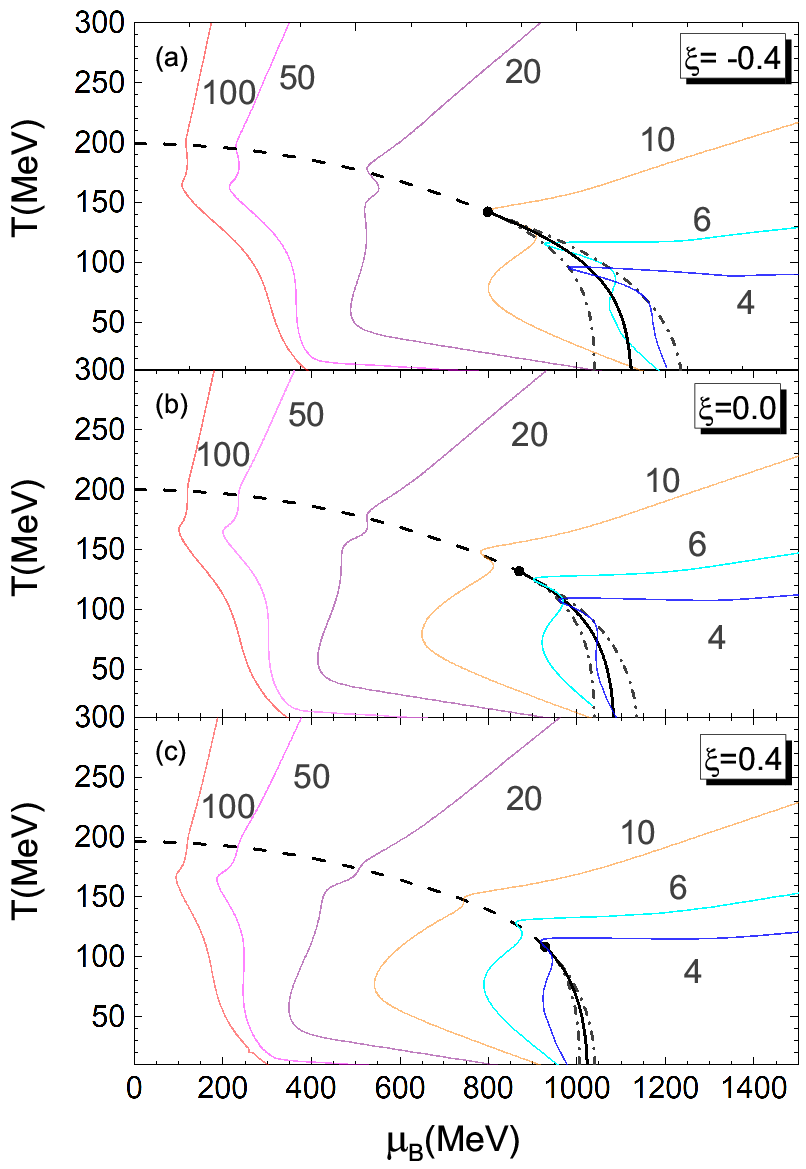}
\end{minipage}
\caption{ QCD phase diagrams for different anisotropic parameter. (a) $\xi=-0.4$ with a stretched momentum distribution along the anisotropic direction (b) $\xi=0$ with a isotropic momentum distribution (c) $\xi$=0.4 with a squeezed momentum distribution along the anisotropic direction. The isentropic trajectories with $s/\rho_B=100, 50, 20, 10, 6, 4$ are also plotted in the three cases.}
\label{fig:4}
\end{figure}

The full QCD phase diagrams are plotted in Fig.~\ref{fig:4}b for the isotropic quark matter~($\xi=0$) and in Fig.~\ref{fig:4}(a) and (c) for the anisotropic quark matter with $\xi=\pm0.4$.
Fig.~\ref{fig:4} indicates that there is a close relationship between the phase structure and the anisotropic parameter $\xi$, in particular, in the area  near the critical region and the first-order phase transition. 
In the case of $\xi=-0.4$ with a stretched momentum distribution along the anisotropic direction, the CEP of chiral phase transition moves to a higher temperature and smaller chemical potential, compared with the isotropic quark matter with $\xi=0$. The associated spinodal region of the first-order phase transition is correspondingly enlarged, which means that the metastable and unstable phases exist in more wider range.  Fig.~\ref{fig:4}(a) also shows that  at lower temperatures the firs-order phase transition has a larger chemical potential than that of $\xi=0$. Compared with the result of $\xi=-0.4$, in the case of $\xi=0.4$ with a squeezed momentum distribution along the anisotropic direction,  the opposite trend appears, and the range of first-order phase transition as well as the associated spinodal structure shrink in the phase diagram.  Additionally, with the inclusion of entanglement interaction or eight-quark interaction, the numerical results indicate that  the QCD phase structure shows a similar trend at low temperature as shown in Fig.~\ref{fig:4}.

\begin{figure*}[htbp]
%\begin{minipage}{\columnwidth}
\begin{center}
\includegraphics[scale=0.5]{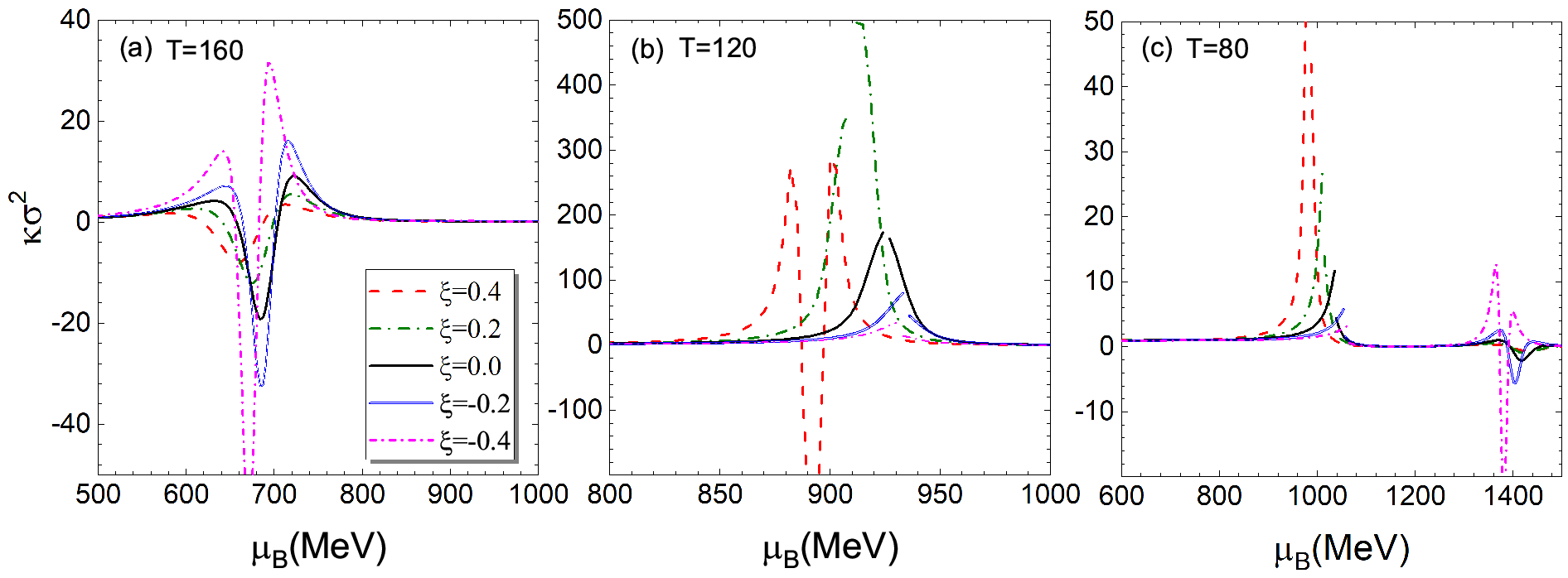}
\caption{ Baryon number kurtosis (a) near the crossover phase transition, (b) near the critical region, (c) near the first-order region}
\label{fig:5}   
\end{center}
%\end{minipage}
\end{figure*}
%\end{widetext}

\begin{figure*}[htbp]
%\begin{minipage}{\columnwidth}
\begin{center}
\includegraphics[scale=0.5]{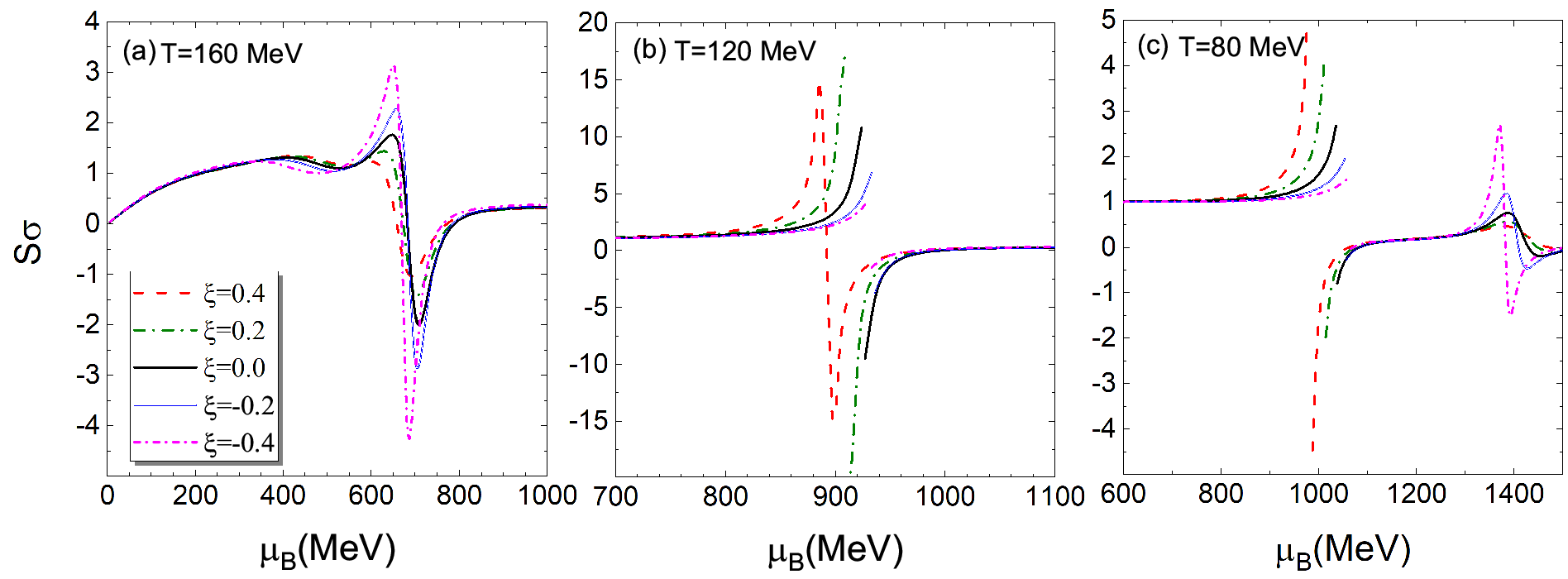}
\caption{ Baryon number skewness (a) near the crossover phase transition, (b) near the critical region, (c) near the first-order region}
\label{fig:6}   
\end{center}
%\end{minipage}
\end{figure*}

The isentropic trajectories are also plotted in Fig.~\ref{fig:4} for different anisotropic parameter $\xi$. It can be seen that the relation between the isentropic trajectories for $s/\rho_B<20$ and the phase structure at high density highly depends on the type of anisotropy~(stretched or squeezed in the momentum space). For example, the trajectory of $s/\rho_B=6$ in the phase diagram  with $\xi=0.4$  in  Fig.~\ref{fig:4}(c) passes  through the chiral crossover transformation line, but the isentropic trajectory with the same $s/\rho_B$ for $\xi=-0.4$  in  Fig.~\ref{fig:4} (a) passes through the first-order phase transition.
Since the entropy per baryon is connected to the collision energy in  HIC experiments, the estimate of the initial entropy density and   entropy density per baryon can be extracted for different center-of-mass energies~\cite{Motta20}.
%Besides, the fluctuations of conserved charges are sensitive to the QCD phase structure. Therefore, the fluctuations of conserved charges are naturally  relevant to the anisotropy of momentum distribution in HIC experiments for given collision energies. 

Because the fluctuations of conserved charges are sensitive to the QCD phase structure, they are naturally  relevant to the anisotropy of momentum distribution in HIC experiments. 
We plot in Figs.~\ref{fig:5} and \ref{fig:6} the net baryon number kurtosis and skewness as functions of temperature and chemical potential for different anisotropic parameter $\xi$. 
Combined with the phase diagram in Fig.~\ref{fig:4}, Fig.~\ref{fig:5}(a) and Fig.~\ref{fig:6}(a) show that, at higher temperatures closed to the chiral crossover, the kurtosis and skewness in the case of $\xi=-0.4$ with a  stretched momentum distribution along the anisotropic direction are stronger than the squeezed ($\xi=0.4$) and  isotropic ($\xi=0$) ones. 
However, Fig.~\ref{fig:5}(b) and Fig.~\ref{fig:6}(b) indicate that with the decrease of temperature the values of  kurtosis and skewness for $\xi=0.4$ are  larger than those for isotropic ($\xi=0$) and stretched momentum distribution ($\xi<0$) at temperatures near the critical region.  Fig.~\ref{fig:5}(c) and Fig.~\ref{fig:6}(c) also indicate the similar result in the first-order phase transition region. The fluctuations at larger chemical potentials are induced by the chiral phase transition of strange quark.
According to the hydrodynamic simulations and HIC experiments, the squeezed momentum distribution along the beam direction ($\xi>0$) is supported. Therefore, compared with the isotropic quark matter, the values of net baryon kurtosis and skewness are possibly enhanced in anisotropic medium with the decrease of collision energies.  The exact results depend on how far the chemical freeze-out line is away from the phase transition line as well as the anisotropic parameter $\xi$.

The comparative study of net baryon number fluctuations in low-energy effective models, lattice QCD, and heavy-ion collision experiments are important in predicting the QCD phase structure at high density. However, when the anisotropic momentum distribution is considered, the accomplishment of this task is beyond reach at present due to the lack of research in lattice QCD and experimental data. Besides, the  relationship between collision centrality, colliding energy, and the anisotropic parameter $\xi$ are still unknown. The release of more BES II data in the future may provide an opportunity to investigate these aspects. When conditions permit, a comparison of the kurtosis and skewness of net baryon number fluctuaions with the anisotropy of quark momentum in this effective model with those of  experimental data and/or lattice QCD  will be conducted. Only some qualitative results are presented in this preliminary investigation.

\section{summary}
In this work, we studied the properties of quark matter with the anisotropic momentum distribution in the PNJL model. We analyzed the features of angular dependence of  average momentum and  quark distribution function for different types of momentum anisotropy in the RS scheme. The numerical results indicate that the squeezed (stretched) momentum distribution along the anisotropic direction can be effectively described with the anisotropic parameter $\xi>0$~($\xi<0$).
The calculation also suggests that the QCD phase structure are closely related to the anisotropic parameter $\xi$, in particular in the high density region.  In the case of stretched momentum distribution with $\xi<0$, the range of first-order phase transition are enlarged and the CEP moves to a lower temperature and  smaller chemical potential. The opposite situation occurs in the case of squeezed momentum distribution with $\xi>0$.
%For the case of squeezed momentum distribution with $\xi>0$, the range of first-order phase transition shrinks and the CEP moves to a  lower temperature and larger chemical potential.

We further calculated the kurtosis and skewness of net baryon number fluctuations for different anisotropic parameter.   Since the squeezed momentum distribution along the beam direction is supported by experimental data and hydrodynamic simulations,  $\xi>0$ is required to describe the HIC experiments. In this case,  the numerical results show that the kurtosis and skewness of net baryon number distribution  at lower collision energies are possibly larger than those of isotropic quark matter. The accuracy  depends on the locations of QCD phase transition and the chemical freeze-out line as well as the real anisotropic parameter $\xi$ in experiments.
A comparative study of fluctuations and correlations of conserved charges in quark models with experimental data and lattice QCD will be performed when conditions permit in the future.

%A further study combined with  experimental data will be performed on this issue.

\begin{acknowledgements} 
This work is supported by the National Natural Science Foundation of China under
Grant No. 11875213.
\end{acknowledgements}

%\end{CJK*}
\end{document}